# Topological effects of phonons in GaN and Al$_x$Ga$_{1-x}$N: A potential perspective for tuning phonon transport

Dao-Sheng Tang, Bing-Yang Cao*

(Key laboratory of thermal science and power engineering of Education of Ministry, Department of Engineering Mechanics, Tsinghua University, Beijing 100084, China)

*Corresponding author, Email: caoby@tsinghua.edu.cn

**Abstract:** Tuning thermal transport in semiconductor nanostructures is of great significance for thermal management in information and power electronics. With excellent transport properties, such as ballistic transport, immunity to point defects and disorders, and forbidden backscattering, topological phonon surface states show remarkable potential in addressing this issue. Herein, topological phonon analyses are performed on hexagonal wurtzite GaN to check the topological characteristics of phonons. And other nitrides of the same family, i.e., AlN and AlGaN alloy, are also calculated from a topological phonon phase transition perspective. With the aid of first-principle calculations and topological phonon theory, Weyl phonon states, which host surfaces states without backscattering, are investigated for all these materials. The results show that there is no nontrivial topological phonon state in GaN. However, by introducing Al atoms, i.e., in wurtzite type AlN and AlGaN, more than one Weyl phonon point is found, confirmed by obvious topological characteristics, including non-zero integer topological charges, source/sink in Berry curvature distributions, surface local density of states and surface arcs. As AlN and AlGaN are typical materials in AlGaN/GaN heterostructure based electronics, the existence of topological phonon states in them will benefit thermal management by facilitating the design of one-way interfacial phonon transport without backscattering.

# I. Introduction

Gallium nitride (GaN) with wurtzite structure, a representative of wide bandgap semiconductors, is currently one of the most important semiconductor materials for information and power electronics [1,2]. With the increase of working power and integration density, near-junction thermal management is becoming more and more important, including accurate thermal simulations, optimization of thermal transport (i.e., phonon transport in dielectric materials), and efficient heat dissipation [3-6]. As a typical GaN-based electronic, the GaN high electron mobility transistor (HEMT) has received much attention in the past decade. Theoretical and experimental analyses show that thermal resistance in HEMTs is mainly concentrated in the near-junction region, i.e., the channel layer. Based on hybrid simulations combining phonon Monte Carlo methods and Fourier's law calculations, a deeper understanding of near-junction thermal transport is realized. Approximation of one-dimensional thermal transport used in many types of research is inappropriate for HEMT thermal analyses. The actual thermal transport process is three-dimensional, where heat is generated by a dot-like heat source and spreads to plane-like areas crossing the channel layer and several interfaces [3]. Hence, thermal transport design methods that can couple with the detailed transport process will be more valuable, while reducing thickness of the channel layer or increasing thermal conductivity of the substrate may be less effective. Particularly, enhancing interfacial phonon thermal transport parallel to the interface is desirable as it can reduce thermal spreading resistance significantly.

Many methods have been proposed to manipulate phonons and facilitate phonon transport [7-19]. The typical regimes include phonon interface/boundary scattering under the phonon ballistic-diffusive regime [7-10], phonon coherence from phonon wave properties [11], surface phonon polarization from phonon-photon coupling [12], phonon localization [13], and phonon response to strain, external fields and phase changes [14-19]. In general, these methods are built based on the current understanding of phonons, which are regarded as both particles and waves, with a Bose-Einstein distribution, yet without an electric charge, spin, or circular polarization. Thus, a deeper understanding of phonon physics will promulgate further developments in materials and transport science. The topological effects of phonons [20-22], which host non-dissipative surface phonon states or without backscattering, promise to provide potential perspectives for tuning phonon transport.

Topological phonon theory has been developed [20-22] inspired by analogy with topological band theory in electrons [23-25] and investigations on the phonon Hall effect [22,26]. Similarly, topological matter physics elucidates a new understanding of electron properties with wave function's topological structure. Phonon physics also benefit significantly from phonon band topology. Besides phonon dispersion relations and scatterings in classical phonon theory, the phonon wave function's topological structures can further divide phonons into different phases; with that, excellent phonon surface states are present in topological nontrivial phonon systems. Following the topological photon (Bosons) research of Haldane et al. [27,28], investigations into the topological effects of phonons in lattice systems with time-reversal symmetry breaking began a decade ago. Zhang et al. [22,26,29] first introduced the topological theory to explain the phonon Hall effect. Liu et al. [20,21] developed systematic topological phonon theory by adopting the physics in electron topological band theory, and proposed possible lattice models with nontrivial topological phonon states (TPSs) as well as effective Hamiltonian models. Similar to the Su-Schrieffer-Heeger (SSH) model for electronic systems, simple phonon topological models were studied in low-dimensional crystal, quasicrystal and amorphous systems [30-33]. Though there are many differences between phonons and electrons, topological theory can be well used in phonons and other boson systems with the aid of lattice symmetry. In recent years, accompanied by advances in semi-metal states in electron systems, the Weyl phonon state, a new topological phonon state, received much attention, and several kinds of Weyl phonon states have been reported in real materials [34-42].

From the perspective of tuning phonon transport, especially phonon transport in specific semiconductor structures, it is critical to realize and apply topological nontrivial phonon states in certain materials. In this work, topological effects of phonons in hexagonal wurtzite GaN and its related materials used in GaN-based HEMTs are analyzed with first-principle calculations and topological phonon theory to check the existence of TPSs, especially Weyl phonon states. It is found that wurtzite AlN and AlGaN contain nontrivial TPSs, specifically Weyl phonons, though the phonon states in GaN are all topologically trivial. Since AlN and AlGaN are both important wide bandgap semiconductors consisting of an AlGaN/GaN heterostructure, the finding of Weyl phonons that host surface phonon states without backscattering promises to be helpful to understand phonon properties and phonon transport tuning in GaN-based electronics.

## II. Topological phonon theory and calculation methods

The first topological state in a time reversal invariant system is the quantum spin Hall insulator [23-25,43]. The time-invariant conjugated double degenerate states are preserved natively by the Kramer theorem since the electron is a spin-1/2 particle. Fu et al. [44] extended this kind of topological state by introducing crystal lattice symmetry. By combining point group symmetry in the lattice and time-reversal symmetry, double degenerate states can also be realized in a spin-less system represented by pseudospin. Since there is no spin for phonons, the most important task to realize the topological nontrivial insulator phonon state is to create pseudospin, which refers to a topological crystalline insulator. Requirements for topological insulator phonon states are demanding since the crystal symmetry used for bulk states should also be present in corresponding surfaces, which is one of the reasons why real materials containing topological insulator phonon states are still not reported in the literature. Topological semi-metals such as Dirac, Weyl, triple degenerate nodal point, and nodal line and ring fermions are new topological states in semi-metal systems [45,46]. The corresponding Weyl points in three-dimensional phonon systems only require double degenerate band crossing in the bulk system without a requirement for surface lattice symmetry, raising more possibilities of TPSs in real lattices [20,34,35,37-42,47]. For phonons in phase with topological insulator states and Weyl states, the topological nontrivial surface states are bound states immune to disorder and point defects (two-dimensional topological insulator states) with forbidden backscattering (three-dimensional topological insulator states and Weyl states), hosting robust one-way propagation characteristics protected by bulk topological characteristics of phonons.

In studies of topological states, the definition of topological quantities is crucial and depends on a detailed understanding of the underlying physics. In the crystal lattice, the lattice vibration of associated atoms is described by a set of dynamic equations based on Newton's law of motion. By adopting the harmonic approximation and lattice wave solutions, an eigenvalue problem can be obtained, i.e., the phonon dynamic equation. Information of harmonic phonons is fully contained in the equation

$$\mathbf{D}(\mathbf{k})\mathbf{u}(\mathbf{k}) = \omega^2(\mathbf{k})\mathbf{u}(\mathbf{k}), \tag{1}$$

where **D** is dynamic matrices, **u** is an eigenstate, and $\omega$ is phonon frequency. They are all functions of the phonon wave vector **k**. While more attention is paid to phonon eigenvalues, i.e., phonon dispersion relations, as mentioned above, much information is hidden in eigenvectors, like the topological structure of eigenvectors. By properly defining a phonon wave function [21], topological quantities used in electron systems can also be defined in phonon systems, where the Berry connection is defined as

$$\mathbf{A}_{n,k} = -i\langle \mathbf{u}_{n,\mathbf{k}} | \nabla_{\mathbf{k}} | \mathbf{u}_{n,\mathbf{k}} \rangle. \tag{2}$$

Then Berry curvature is defined as $\mathbf{\Omega}_{n,\mathbf{k}} = \nabla_{\mathbf{k}} \times \mathbf{A}_{n,\mathbf{k}}$, where $n$ is the band index, and the Berry phase is defined as integral of Berry connection along a path or loop

$$\gamma_n = \int_L \mathbf{A} \cdot d\mathbf{k}. \tag{3}$$

Topological invariant is the basic physical quantity to identify the topological properties of the system directly. For the Weyl points focused on in this work, the topological invariant is the topological charge, i.e., the Chern number. The three-dimensional condition is defined in a closed surface containing the Weyl point as

$$C = \frac{1}{2\pi} \oint \mathbf{\Omega}_{n,\mathbf{k}} d^2\mathbf{k}. \tag{4}$$

Currently there are mainly two kinds of Weyl points, including the single Weyl point with Chern number equal to $\pm 1$, and the double Weyl point with Chern number equal to $\pm 2$. Most recently, Weyl points beyond these two kinds have also been reported [48]. Weyl point characteristics can be illustrated by the evolution of the Wannier charge center, the distribution of the Berry curvature, and surface states, besides topological charge calculations. At the Weyl point, a source or sink can be observed in the Berry curvature distribution in a plane containing this Weyl point. Surface states are direct evidence of nontrivial TPSs, in which the surface arc is the characteristics of Weyl points. In detail, the iterative Green's function method implemented in open-source software WannierTools [49] is adopted to calculate the surface local density of states (LDOS) with a phonon tight-binding (TB) Hamiltonian from second-order interatomic force constants (IFCs) and the Wilson-loop method is used to calculate Wannier charge center evolutions. The basic quantities, including second-order IFCs and phonon eigenvectors, are calculated by adopting the frozen phonon method from first principles accompanied the open-source software Phonopy [50]. All first principle calculations are performed within the framework of density functional theory as implemented in the Vienna Ab-initio Simulation Package (VASP) [51]. Projective augmented wave pseudopotential [52], and the generalized gradient approximation in the Perdew-Burke-Ernzerhof form [53] for the exchange-correlation functional are adopted. The kinetic energy cutoff for a plane wave basis of 800 eV is employed, and the Brillouin zone is sampled using converged 12×12×9 Gamma-centered $k$-mesh grids. The conventional unit cells are relaxed until the residual stress and the maximum forces acting on each atom are smaller than $10^{-2}$ kbar and $10^{-7}$ eV/Å. The supercells for wurtzite GaN and AlN are both built on conventional unit cells with size 4×4×3 and 4×4×4, respectively. For AlGaN, the size depends on the number of atoms in the unit cell, where a 3×3×3 supercell is used for an 8-atom unit cell, and a 4×4×4 supercell is used for 4-atom unit cell. Born effective charges and dielectric constants are calculated using density functional perturbation theory to consider the effects of the long-range coulomb interaction on second-order IFCs, which is often called non-analytical correction (NAC) and mainly affects phonon states near the Brillouin zone center, i.e., the Gamma point.

III. Results and discussion

A. Topological effects of phonons in GaN

Wurtzite structure is the most common lattice structure for GaN as it is thermodynamically stable. Illustrated by Figure 1 (a), the wurtzite structure is a hexagonal structure with lattice constants *a* and *c*, and internal parameter *u*, belonging to space group P6$_3$mc (No. 186). The wurtzite structure host twelve lattice symmetry operations, including 3-fold rotation symmetry, 2-fold, and 6-fold non-symmorphic screw rotation symmetry, mirror symmetry, and screw mirror symmetry. The Brillouin zone is also shown in Figure 1 (b), where high symmetry points and paths are labeled.

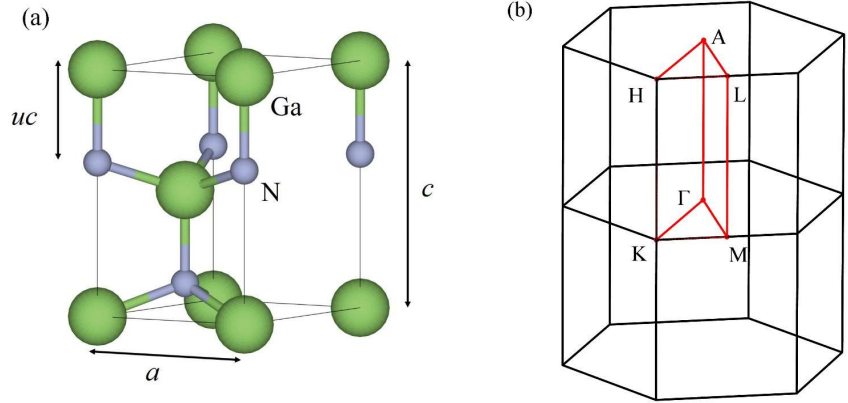

Figure 1. (a) Lattice structure and (b) Brillouin zone of wurtzite GaN. The letters *a* and *c* in (a) indicate lattice constants and *u* internal parameter, and letters in (b) show high symmetry points.

As mentioned above, it is hard to satisfy the requirements for the topological insulator states of phonons, as shown in the wurtzite structure in the following discussion. Also, surface phonons of nodal line (ring) states do not contain excellent transport properties such as forbidden backscattering. Thus, Weyl point states are focused on in this work. Detecting topological nontrivial phonon states starts from analyses of phonon dispersions, especially phonon degeneracy at specific points, lines, and planes. The phonon dispersions of GaN are shown in Figure 2. It is noted ahead of the detailed discussion that NAC is still not properly included in the phonon TB Hamiltonian for materials with strong polarization, which is under development [49]. Thus, phonon dispersion relations without NAC are used in this work. And phonon bands near the Gamma point are then not considered in the topological analyses. As seen from Figure 2, the circled phonon band crossing points (BCPs) are not affected by NAC for slight differences in the magnitude of phonon frequencies. For all calculations in this work, we have confirmed the BCPs discussed are not affected by NAC, except the last example of $Al_3GaN_4$. Also, including NAC will not bring in new BCPs in general since it will introduce LO-TO splitting at the Gamma point. Turning attention to phonon dispersion of GaN, due to the large mass ratio between Ga and N atoms, a large gap exists in phonon dispersion relations, which may reduce many BCPs.

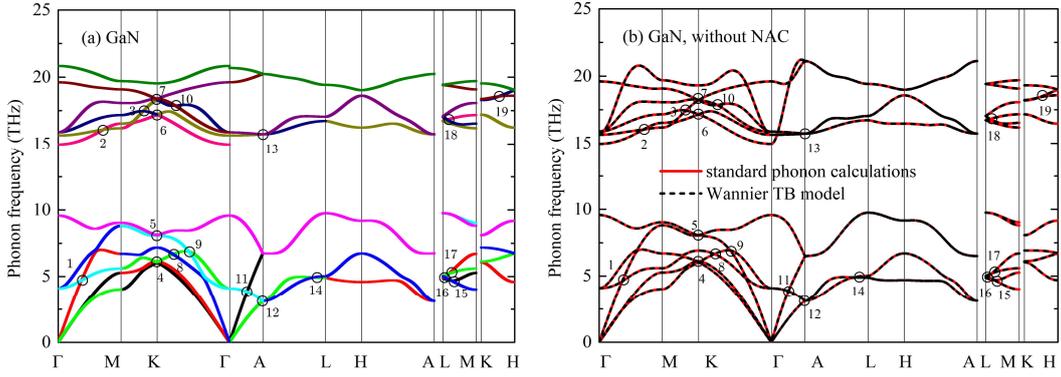

Figure 2. Phonon dispersions of wurtzite GaN (a) with NAC from standard phonon calculations and (b) without NAC from standard phonon calculations and Wannier TB model Hamiltonian. BCPs are labeled with black circles in both (a) and (b).

For topological insulator states, two pairs of phonon degenerate phonon modes are needed at the same wave vector in both the bulk and surface Brillouin zones. However, only 6-fold screw rotation symmetry exists, instead of 6-fold rotation symmetry, which cannot support topological insulator states. To check Weyl phonon points in GaN, all phonon BCPs at high symmetry paths and planes (including planes $k_z=0$, $k_z=0.5$, $k_x=0$, and $k_y=0$ in Cartesian coordinates) in the Brillouin zone are searched by calculating topological charges and Berry curvature distribution. Since no topological characteristics of phonons is found after these calculations, it is concluded that wurtzite GaN is a trivial topological phonon system. Herein, taking the 14th BCP as an example, trivial surface states at the (0001) surface are presented and provide an intuitive understanding of the calculations, which can also be regarded as comparisons with the following nontrivial results. In electronic components, wurtzite nitrides are generally grown along the polar axis, i.e. the (0001) direction. Therefore, the (0001) surface is selected in this work for surface state calculations. As seen from Figure 3, in the projection of the iso-frequency plane on the (0001) surface, instead of a surface arc of Weyl points, there are only uninterrupted annular shapes. The LDOS of the surface along high symmetry paths in the surface Brillouin zone is shown in Figure 4. The surface states are calculated by both the iterative Green's function method and directly solving the eigenstates equation of phonons of a slab structure. Both results are consistent. It is noted here that there are two kinds of (0001) surfaces, including a Ga surface and an N surface, which are equivalent to bottom and top surfaces, respectively. In the following discussions, the surfaces with significant topological characteristics are focused on while surface type is not distinguished particularly.

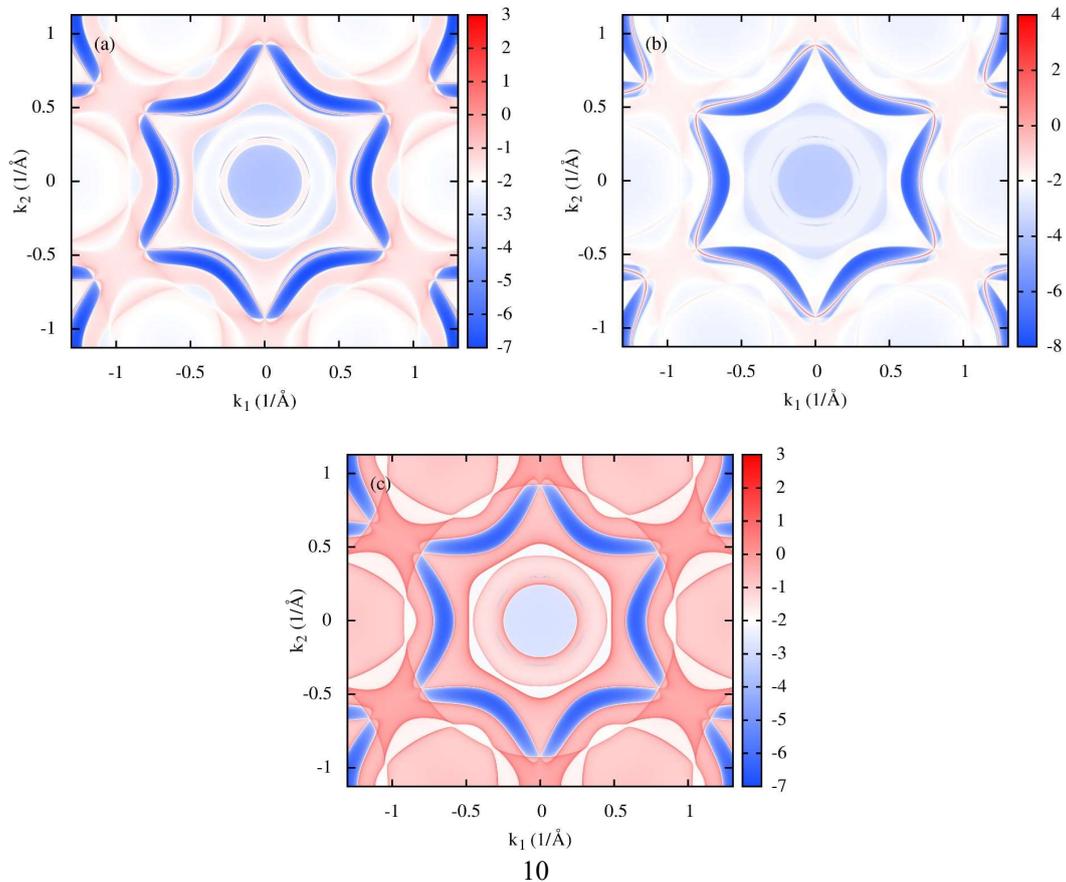

Figure 3. Projections of the iso-frequency plane at 4.93 THz of (a) top surface states and (b) bottom surface states and (c) bulk states on the (0001) surface. Closed surface states can be observed in (a) and (b).

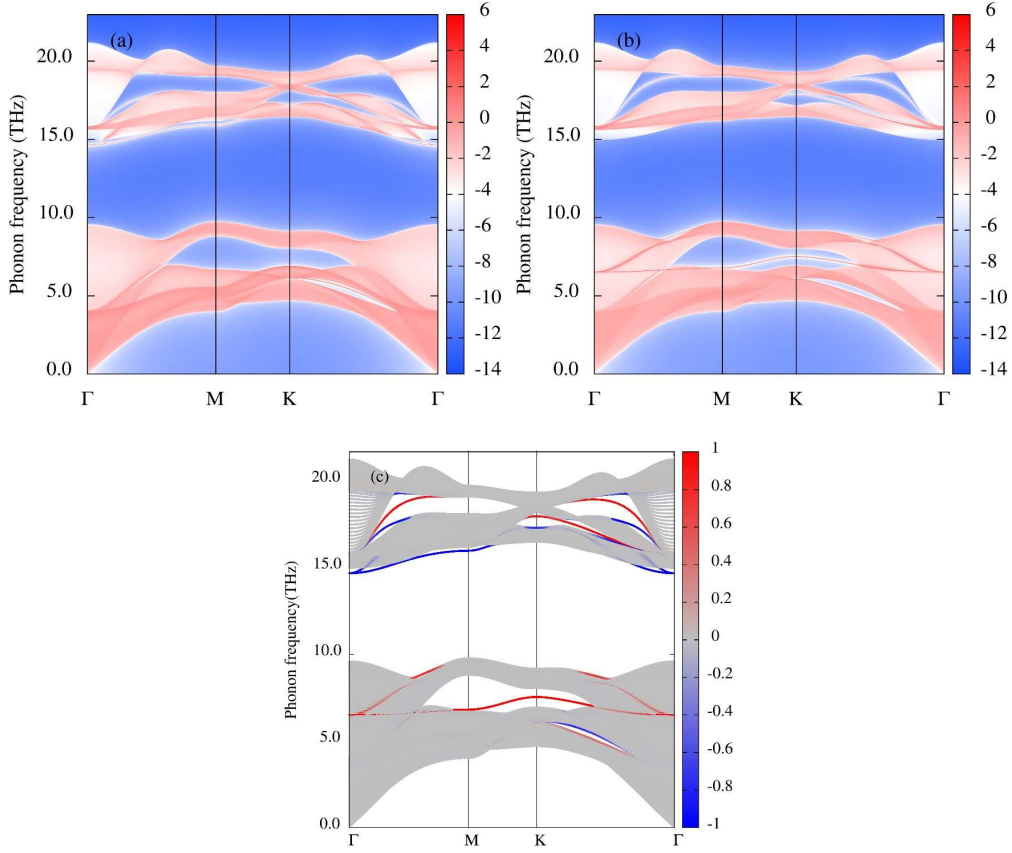

Figure 4. The LDOS for the (0001) surface along the high symmetry paths in surface Brillouin zone calculated by the iterative Green's function method in (a), (b) and directly solving the eigenstate equation of phonon of a slab structure in (c). The blue and red lines indicate the surface states on the top and bottom surfaces of a slab structure, consistent with those illustrated in (a) and (b).

### B. Topological phonon phase transition by introducing Al

In References [35] and [54], conclusions and inferences show that the Weyl complex and type II Weyl phonons can exist in the wurtzite structure (Space group P6$_3$mc, No. 186) at the high symmetry points Gamma (A) and K (H) [35] and the high symmetry plane instead of the high symmetry path [54]. Hence, topological phonon phase transitions may be realized in GaN of wurtzite structure as we have verified that there is no Weyl point in this system.

The classical model in topological phonon theory, i.e., SSH-like model for phonons [31,32] and the effective Hamiltonian model for topological insulator phonon states [20], indicate that second-order IFCs are critical parameters in determining topological properties of the phonon system and the requirement for lattice symmetry is necessary but not sufficient.

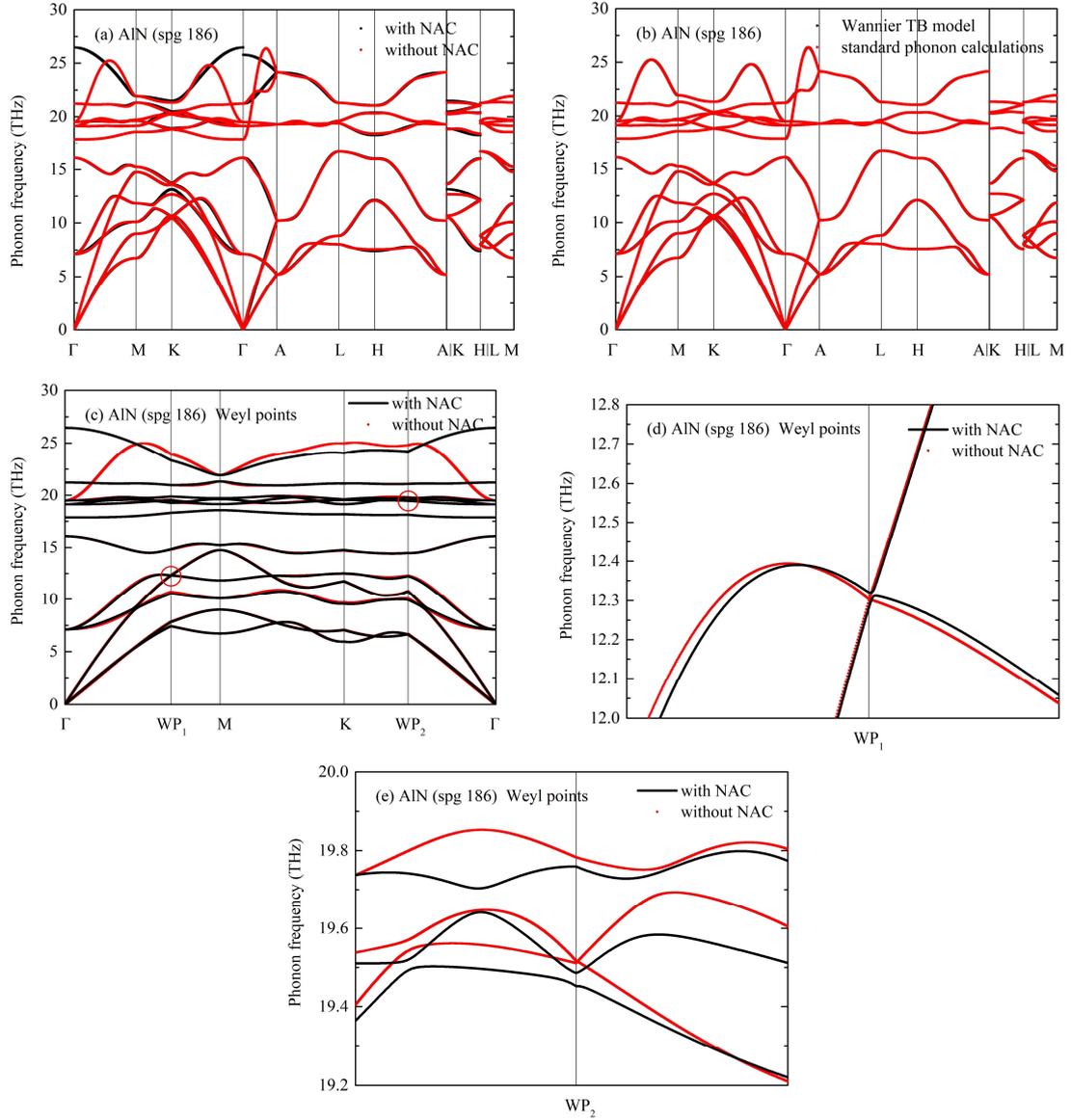

Figure 5. Phonon dispersion of wurtzite AlN. (a) The phonon dispersion of AlN with and without NAC. (b) The phonon dispersion obtained by standard phonon calculations and reproduced by Wannier tight-binding model Hamiltonian. (c) The phonon dispersion along the path including two different Weyl points marked as $WP_1$ and $WP_2$, is not affected by NAC. (d) and (e) are enlarged views of two Weyl points in (c).

In this section, we further check GaN related structures, including wurtzite GaN with tensile strain along and perpendicular to polar axis, III-A nitrides (AlN and BN), and the most common GaN alloys in electronics, i.e. AlGaN. It is found that single Weyl phonons exist in AlN and four types of AlGaN when Al atoms are introduced into GaN, while no Weyl phonon is found in other systems. Particularly, discussions on wurtzite AlN and wurtzite-like AlGaN (space group P3m1, No. 156) are more valuable since these two materials are used in real electronics. The wurtzite-like AlGaN can be formed by replacing one Ga atom with one Al atom in wurtzite GaN. Therefore, in the followings, we focus on wurtzite AlN and wurtzite-like AlGaN. For other AlGaN systems, we only introduce $Al_3GaN_4$ (space group P-43m, No. 215) as an example.

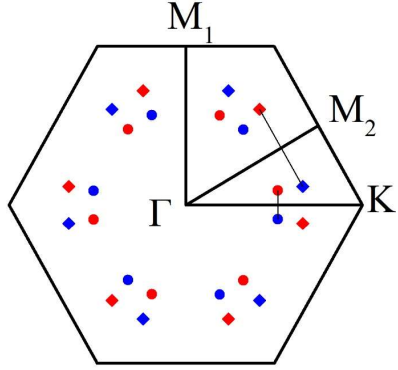

Figure 6. Weyl points in Brillouin zone at plane $k_z$=0 shown by the black hexagon. The dots with different shapes represent different Weyl points, where the rhombic dots represent Weyl point 1 and its equivalent points as shown in Figure 5 (c), and the circle dots represent Weyl point 2 and its equivalent points. The dots in red mean Weyl points with positive chirality equal to 1, while those in blue mean Weyl points with negative chirality equal to -1. Weyl points linked by lines represent a couple of Weyl points.

The phonon dispersions of AlN are shown in Figure 5, where phonon dispersions of AlN along high symmetry paths are shown in Figures 5 (a) and (b). In Figures 5 (c), (d) and (e), nontrivial band crossing points with and without considering NAC are both presented, and the figures show the crossing points are not affected by NAC. Slight frequency changes may occur, resulting in a small displacement of the Weyl point with NAC, explaining the slight gap in the NAC case. The systematic topological phonon analyses here are the same as those in the last section. The positions of Weyl phonons in the Brillouin zone are labeled in Figure 6. Nontrivial topological charges are calculated using the Wilson-loop method and confirmed by Wannier charge center evolution (Figure 7). Besides, the Berry curvature distribution where source and sink are present at Weyl points with positive and negative chirality is shown in Figure 8, in which one source and one sink are marked with red and blue dots, respectively. Surface arcs in the projection of the iso-frequency plane on the (0001) surfaces (Figure 9), and topological surface states illustrated by surface LDOS at specific paths linking two Weyl phonons with opposite chirality (Figure 10) provide solid evidence for nontrivial TPSs. Though there are rich BCPs along the high symmetry paths, no Weyl point is found in them. Two different Weyl points (equivalent points are not counted here) in AlN exist in the high symmetry plane, the $k_x$-$k_y$ plane with $k_z$=0. For the first kind of Weyl phonon, the points with positive chirality 1 and negative chirality -1 are (0.2989, 0.1249, 0.0000) and (0.4239, -0.1246, 0.0000) in fractional coordinates, and their respective equivalent points, which are all shown in Figure 6. The 4th and 5th bands cross at this point with frequency 12.23 THz. For the second kind of Weyl point, the points with chirality 1 and -1 are (0.3549, 0.1169, 0.0000) and (0.3549, -0.2380, 0.0000), and their equivalent points in Brillouin zone based on lattice symmetry. These Weyl points are formed by the crossing of the 8th and 9th bands at frequency 19.57 THz. In Figures 9 (b) and (d), the projection of the iso-frequency plane at the (0001) surface is enlarged to clearly show the surface arcs linking a couple of Weyl phonons. The surface states for the second kind of Weyl point are not as clean as those in the first kind, as there are also trivial surface states near this point with a frequency around 19.57 THz, i.e., the lower single line below 19.6 THz in Figure 10 (b).

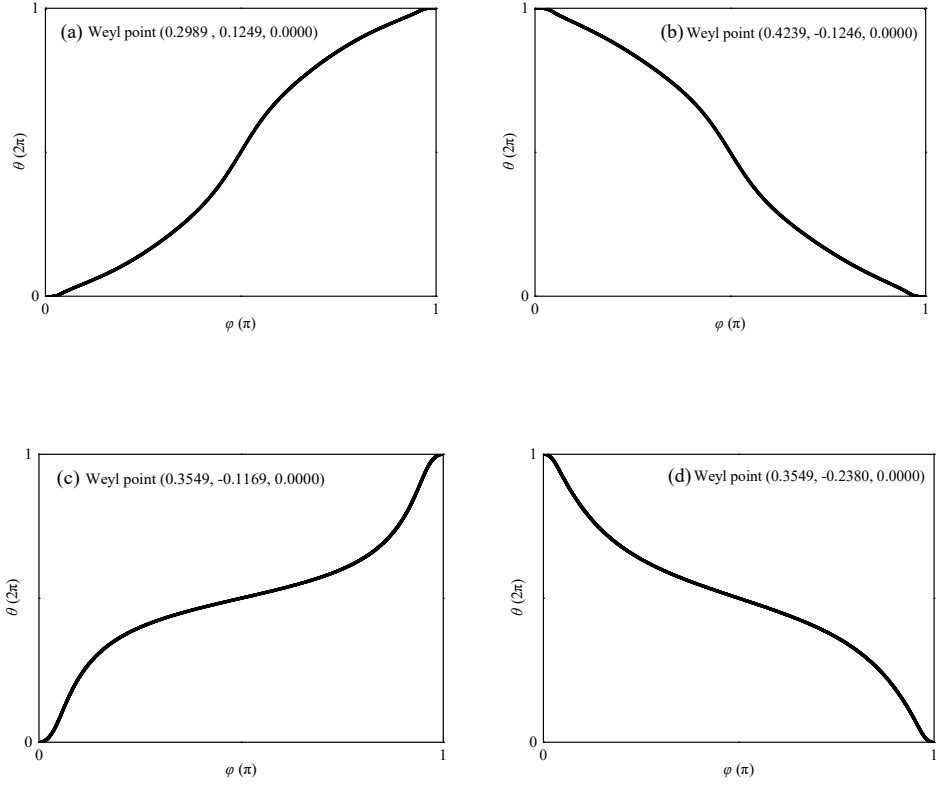

Figure 7. (a), (b) The Wannier charge center evolution for a couple Weyl points (0.2989, 0.1249, 0.0000) and (0.4239, -0.1246, 0.0000), with positive chirality 1, and negative chirality -1, respectively. (c), (d) The Wannier charge center evolution for a couple Weyl points (0.3549, 0.1169, 0.0000) and (0.3549, -0.2380, 0.0000), with positive chirality 1, and negative chirality -1, respectively.

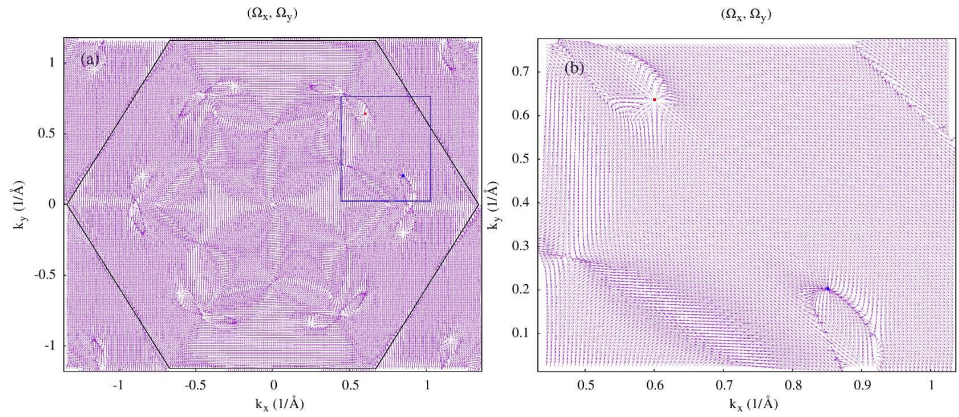

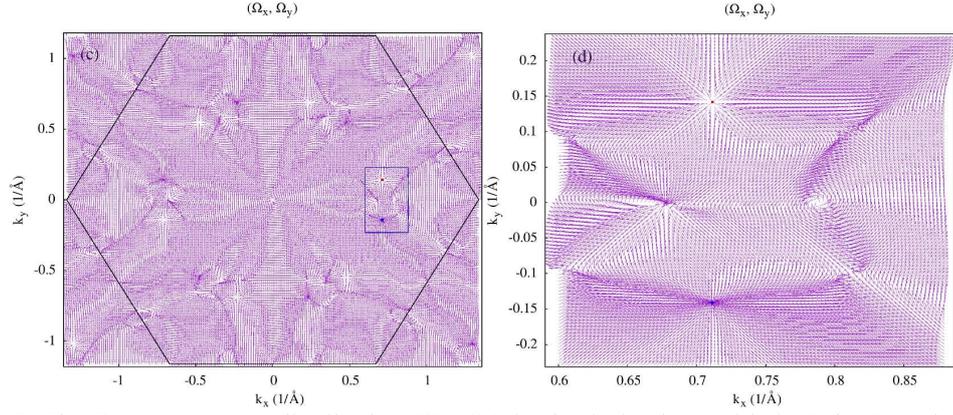

Figure 8. The Berry curvature distribution ($\Omega_x$, $\Omega_y$) in the $k_x$-$k_y$ plane with $k_z$=0 for (a) 4th, (c) 8th phonon band. (b) and (d) are enlarged views of the zones encircled by blue rectangles in (a) and (c), respectively. The source and sink can be found from the distributions. The red and blue dots represent the source and sink of Berry curvature distribution, illustrating positive and negative chirality. The black hexagon represents the $k_z$=0 plane of the Brillouin zone of wurtzite AlN.

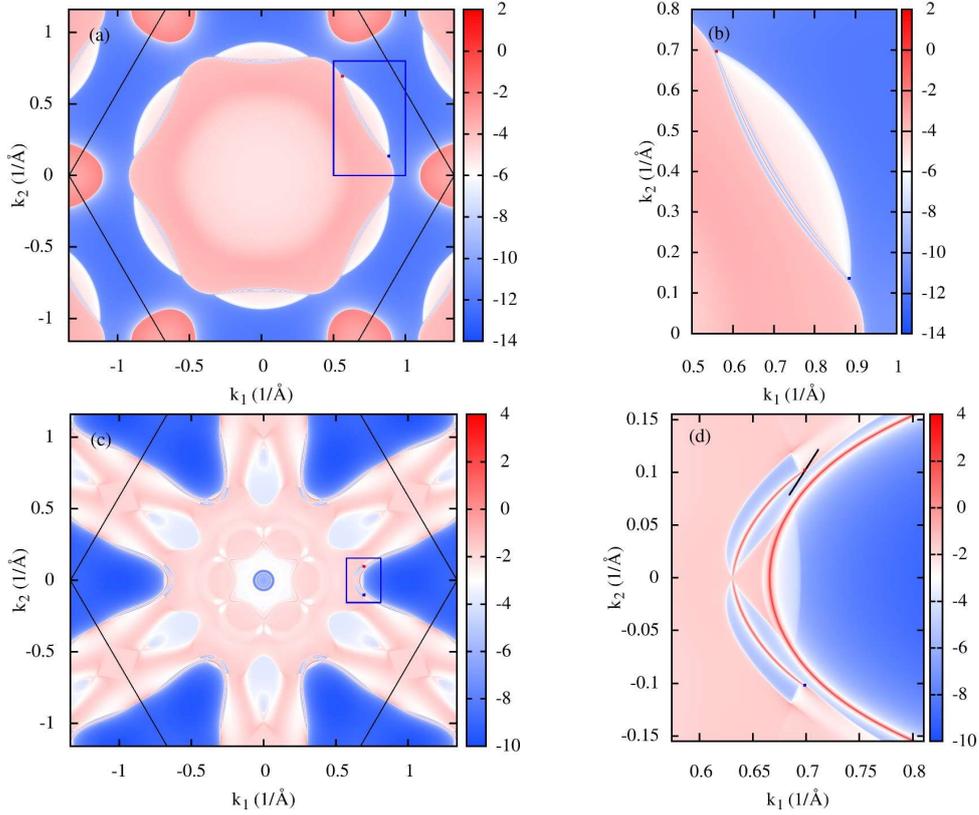

Figure 9. Weyl surface arcs. The projection of the iso-frequency plane of surface states on the (0001) surface Brillouin zone at a frequency equal to (a) 12.23 THz and (b) 19.57 THz. The red and blue dots represent Weyl points with positive chirality 1 and negative chirality -1, respectively, which are linked by phonon surface arcs. The black hexagon indicates the surface Brillouin zone of the (0001) surface

in wurtzite AlN. (b) and (d) are enlarged views of the zones encircled by blue rectangles in (a) and (c), respectively.

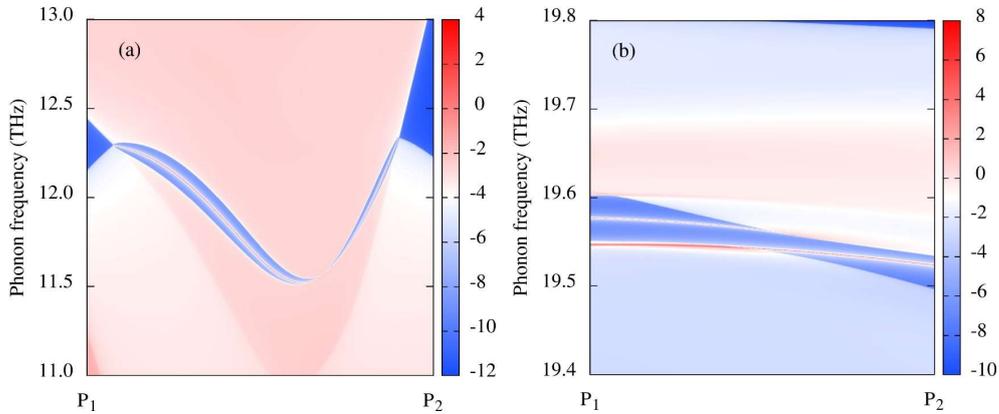

Figure 10. The LDOS for the (0001) surface along the selected path (a) linked a couple of Weyl points (b) across a Weyl point. In (a), the single line between a couple of Weyl points indicates the topological surface states, where $P_1$ and $P_2$ are on the extension line connecting two Weyl points. In (b), the top single line below frequency 19.6 THz indicates the topological surface states, where $P_1$ and $P_2$ are illustrated in Figure 9 with a black line segment.

$Al_xGa_{1-x}N$ is the most common alloy of GaN, an important material in the AlGaN/GaN heterostructure. Experimentally, the proportions of Al and Ga atoms can be controlled. However, it may not be easy to calculate phonon properties from first principles for structures with a large unit cell. In the public material database, there are seven types of AlGaN structures, including structures with space group P3m1 (No. 156), P-43m (No. 215), P-4m2 (No. 115), and three other structures with very low symmetry. $AlGaN_2$ with space group P3m1 is a structure corresponding to wurtzite GaN and more similar to $Al_xGa_{1-x}N$ used in AlGaN/GaN heterostructure. Consequently, the topological phonon properties of this AlGaN alloy are discussed here.

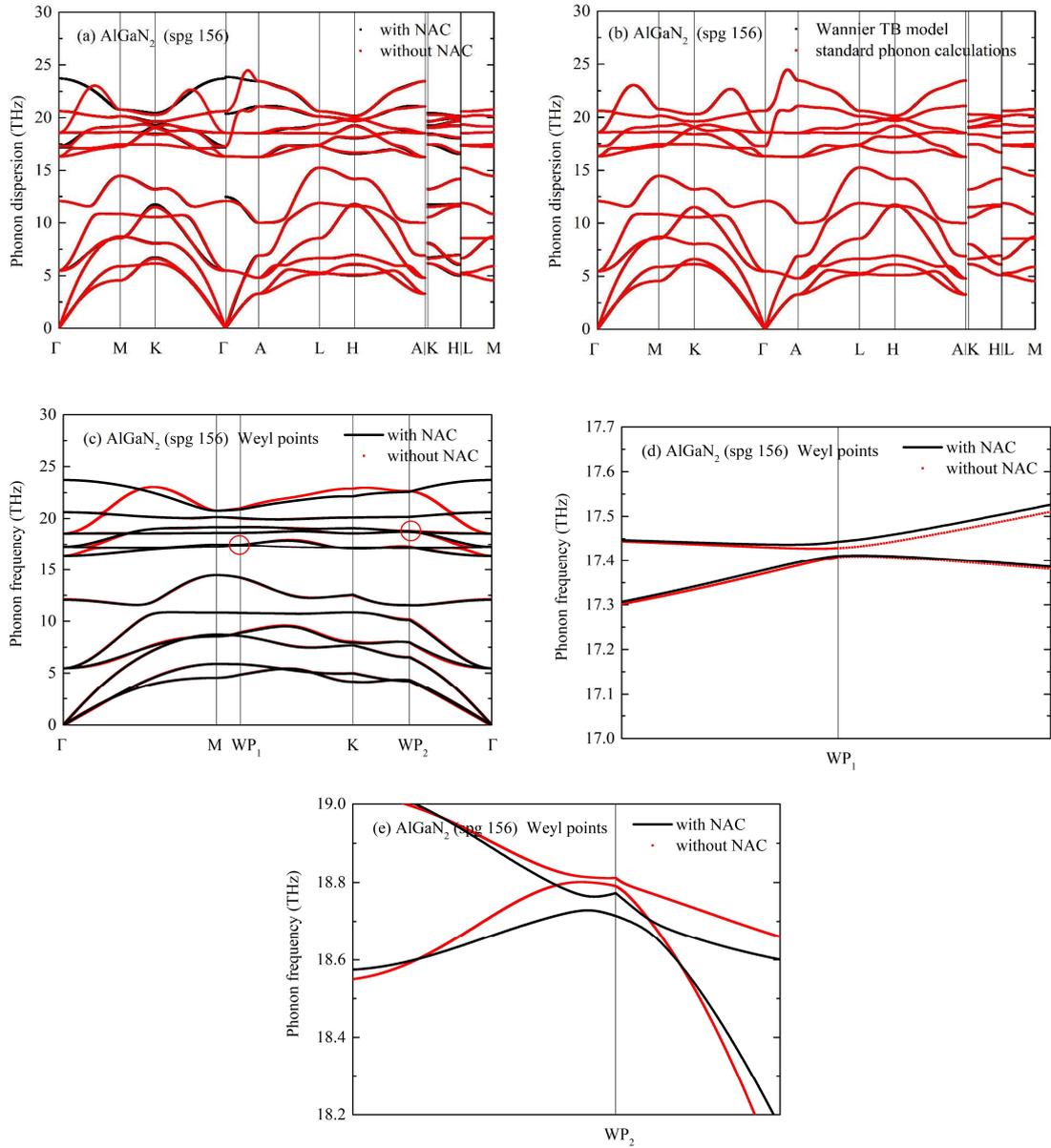

Figure 11. Phonon dispersion of wurtzite AlGaN$_2$. (a) The phonon dispersion of AlGaN$_2$ with and without NAC. (b) The phonon dispersion obtained by standard phonon calculations and reproduced by Wannier tight-binding model Hamiltonian. (c) The phonon dispersion along the path including two different Weyl points marked as WP$_1$ and WP$_2$, is not affected by NAC. (d) and (e) are enlarged views of two Weyl points in (c), where the slight gap in the case without NAC results from the numerical errors and that the Weyl points are difficult to locate vey accurately since they are not at high symmetry paths.

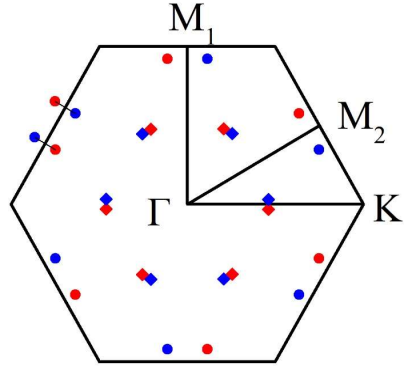

Figure 12. Weyl points in Brillouin zone at plane $k_z$=0 shown by the black hexagon. The dots with different shapes represent different Weyl points, where the rhombic dots represent Weyl point 1, and its equivalent points as shown in Figure 11 (c), and the circle dots represent Weyl point 2 and its equivalent points. The dots in red are Weyl points with positive chirality 1, while those in blue are Weyl points with negative chirality -1. Weyl point represented by the circle dot is coupled with the point linked by the black line outside the first Brillouin zone. Weyl point represented by the rhombic dot is coupled with the nearest one.

Phonon dispersions of AlGaN$_2$ are shown in Figure 11, where phonon dispersions along high symmetry paths are shown in Figures 11 (a) and (b). There are twelve bands since there are four atoms in each unit cell. The calculations show that there are also two different single Weyl points in AlGaN$_2$, which are located at a high symmetry plane, the $k_x$-$k_y$ plane with $k_z$=0. The phonon dispersions along the path across the two kinds of Weyl points are also shown in Figure 11 (c). The first kind of Weyl point is formed by the 7th and 8th bands at (0.4225, 0.0776, 0.0000) in fractional coordinates in Brillouin zone at frequency 17.41 THz. And the second one is formed by the crossing of the 9th and 10th bands at (0.3097, 0.1668, 0.0000) at frequency 18.79 THz. The positions of Weyl points in the Brillouin zone are illustrated in Figure 12. And nontrivial Wannier charge center evolutions and Berry curvature distribution are illustrated by Figures 13 and 14. Unlike Weyl points in AlN, a Weyl point of the first kind is linked by a surface arc with the other one outside the first Brillouin zone, as shown in both Figure 12 and Figures 15 (a) and (b). Actually, the equivalence of Weyl points in AlGaN$_2$ is conserved by lattice symmetry and time reversal invariance symmetry. In Figure 15, distinct surface arcs between Weyl points are illustrated for both kinds of Weyl phonons. Also, both the surface states are distinguishable as that there is no other surface state near frequency 17.41 THz and 18.79 THz (Figure 16).

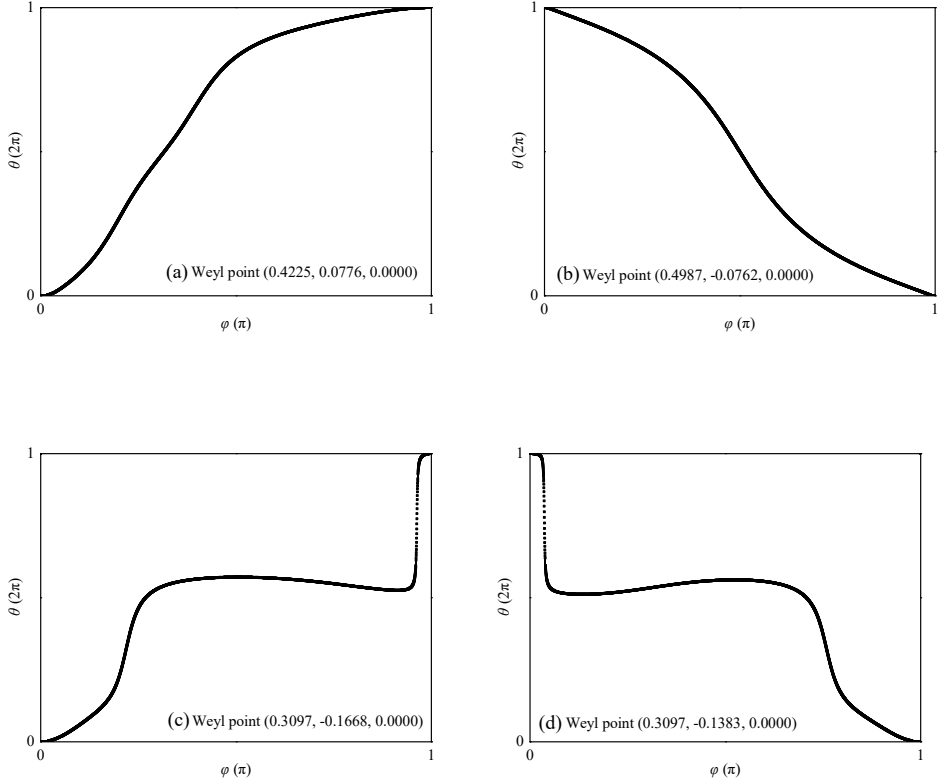

Figure 13. (a), (b) The Wannier charge center evolution for a couple Weyl points (0.4225, 0.0776, 0.0000) and (0.4987, -0.0762, 0.0000), with positive chirality 1, and negative chirality -1, respectively. (c), (d) The Wannier charge center evolution for a couple Weyl points (0.3097, 0.1668, 0.0000) and (0.3097, -0.1383, 0.0000), with positive chirality 1, and negative chirality -1, respectively.

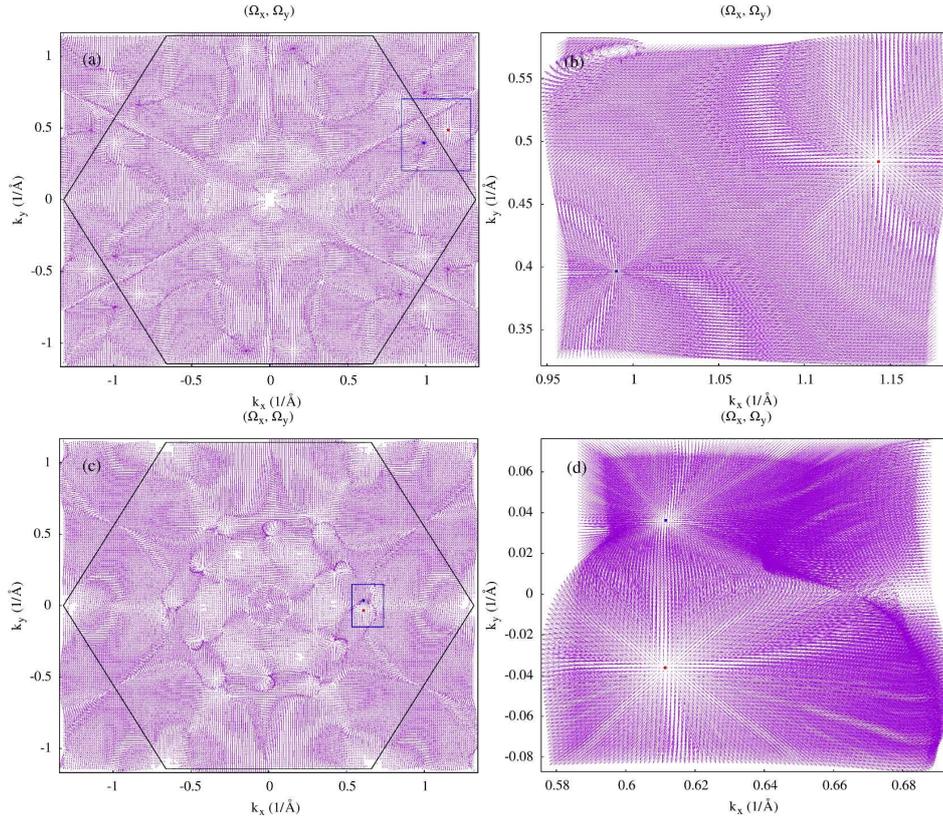

Figure 14. The Berry curvature distribution ($\Omega_x$, $\Omega_y$) in the $k_x$-$k_y$ plane with $k_z=0$ for (a) 7th, (c) 9th phonon band. (b) and (d) are enlarged views of the zones encircled by blue rectangles in (a) and (c), respectively. The source and sink can be found from the distributions. The red and blue dots represent the source and sink of Berry curvature distribution, illustrating positive and negative chirality. The black hexagon represents the $k_z=0$ plane of the Brillouin zone of wurtzite AlGaN$_2$.

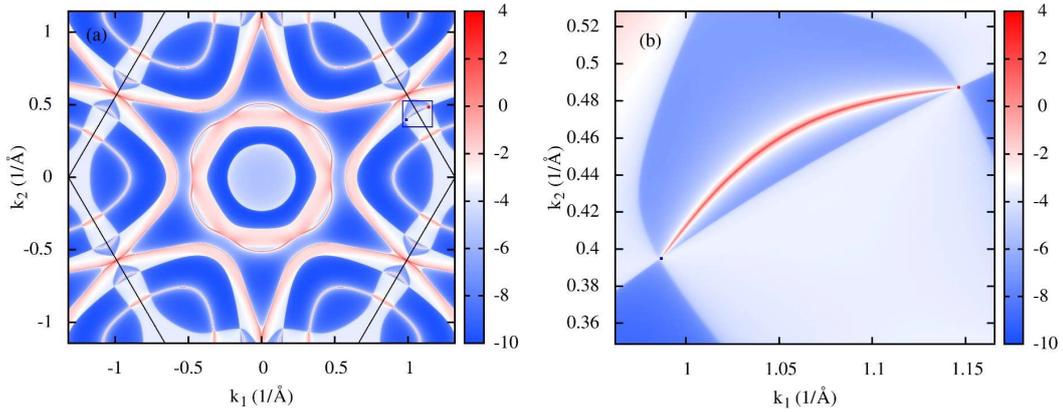

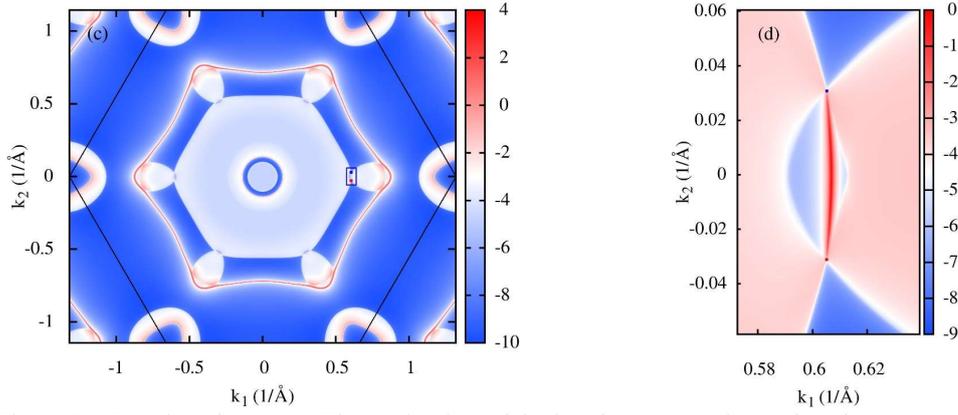

Figure 15. Weyl surface arcs. The projection of the iso-frequency plane of surface states on the (0001) surface at the frequency (a) 17.41 THz and (b) 18.79 THz. The red and blue dots represent Weyl points with positive chirality 1 and negative chirality -1, respectively, which are connected by phonon surface arc. The black hexagon indicates the surface Brillouin zone of the (0001) surface in wurtzite $AlGaN_2$. (b) and (d) are enlarged views of the zones encircled by blue rectangles in (a) and (c), respectively.

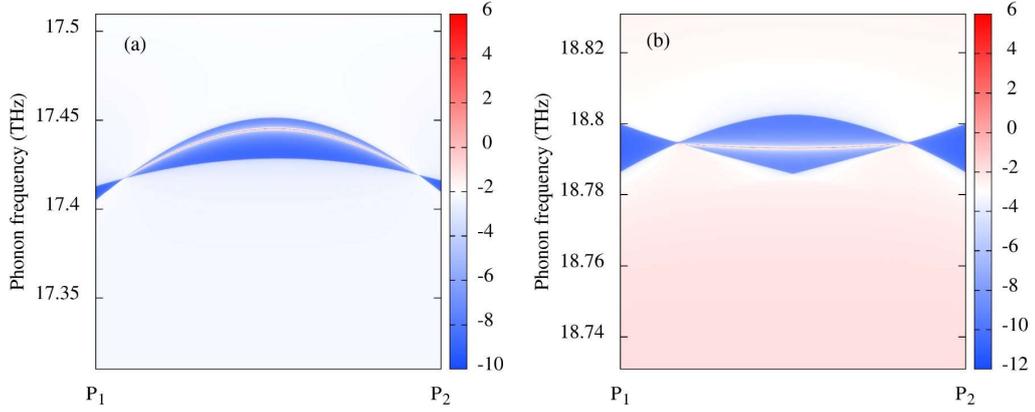

Figure 16. The LDOS for the (0001) surface along the selected path linked a couple of Weyl points. The single line between a couple of Weyl points indicates the topological surface states, where $P_1$ and $P_2$ are on the extension line connecting two Weyl points.

Besides in $AlGaN_2$ with space group P3m1, Weyl points are also detected in other AlGaN structures. However, these structures may not be practical for applications in electronics. Therefore, one Weyl point in $Al_3GaN_4$ with space group P-43m (No. 215) is introduced here as a representative. $Al_3GaN_4$ is a cubic structure, where eight atoms including 3 Al, 1 Ga, and 4 N consist of one primitive unit-cell. This structure is similar to zincblende structure, as 3 Al atoms replace 3 Ga atoms in zincblende GaN conventional unit cell. The Brillouin zone Al$_3$GaN$_4$ is also cubic. The phonon dispersions along high symmetry paths are shown in Figure 17.

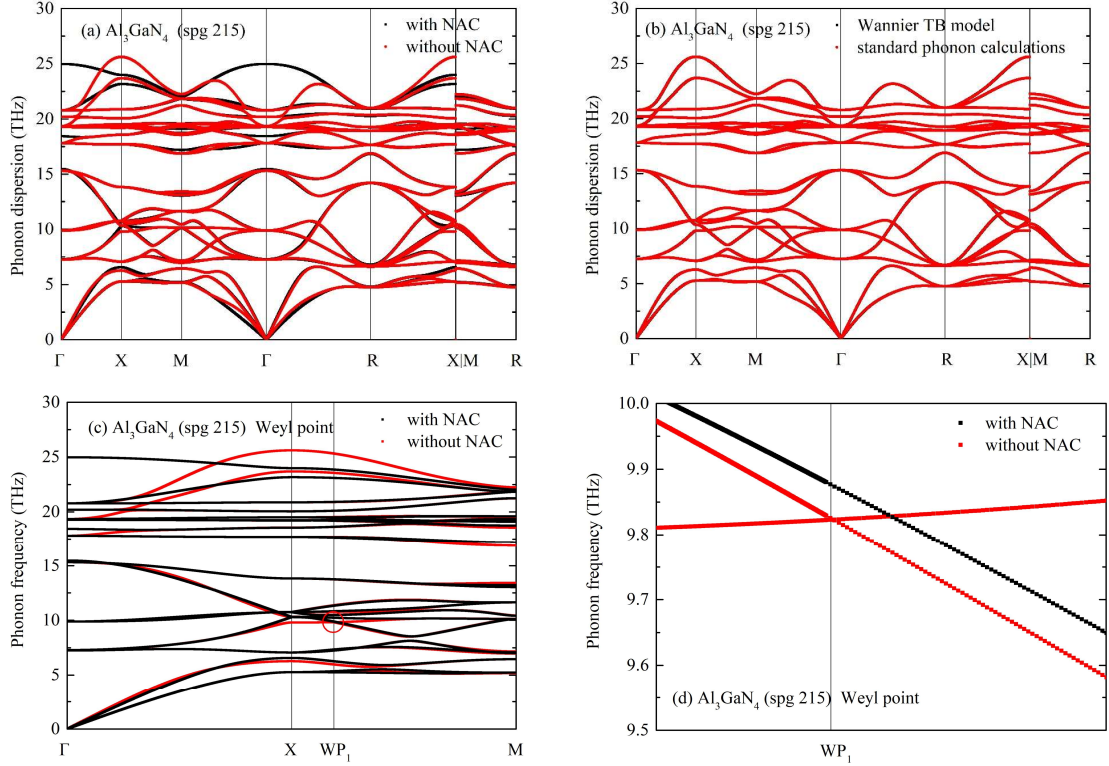

Figure 17. Phonon dispersion of wurtzite Al$_3$GaN$_4$. (a) The phonon dispersion of Al$_3$GaN$_4$ with and without NAC. (b) The phonon dispersion obtained by standard phonon calculations and reproduced by Wannier tight-binding model Hamiltonian. (c) The phonon dispersion along the path including the Weyl point marked as WP$_1$, is affected by NAC. (d) the enlarged view of the Weyl point in (c), which indicates that the Weyl point disappears with NAC.

The calculations show that several single Weyl phonons exist in the $k_z$=0 and $k_z$=0.5 planes of the Brillouin zone. The Weyl point discussed here is formed by the crossing of the 6th and 7th bands at wave vector (0.0941, -0.5000, 0.0000) at a frequency 9.82 THz, located at the high symmetry path X-M, as illustrated in Figure 18. The positions of the Weyl point and its equivalent point are shown in Figure 18 in the $k_z$=0 plane. And characteristics of the Weyl phonons are first verified by nontrivial Wannier charge center evolutions and Berry curvature distribution, as illustrated in Figures 19 and 20. It is noted here this kind of Weyl point also exists in the $k_x$=0 and $k_y$=0 planes based on the lattice symmetry. In Figures 21 and 22, surface arcs and topological surface states can be observed where the surface arcs are relatively long since Weyl points with opposite chirality are far apart. However, surface states in this structure on the (0001) surface are ample, resulting in the overlap of trivial and nontrivial topological surface states. Particularly, it is noted that this kind of Weyl point in $Al_3GaN_4$ may not exist in reality, since the Weyl points disappear after considering NAC, which is clearly shown in Figure 17 (d). After including NAC in phonon dispersion calculations, the band crossing point in the corresponding position disappears. Though generally NAC mainly affects the phonon dispersion near the Brillouin zone, it also affects the phonon dispersion at the Brillouin zone boundary for $Al_3GaN_4$. As seen from Figure 17 (c), this band crossing point in the case without NAC may move to high symmetry point X after considering NAC, and the Weyl phonon state may still be preserved, which can be confirmed when calculation methods are available for cases with NAC.

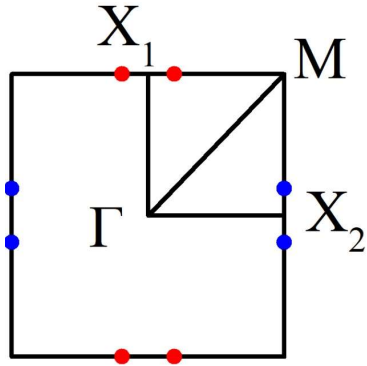

Figure 18. Weyl points in the Brillouin zone at plane $k_z=0$ are shown by the black rectangle. The dots represent the Weyl point shown in Figure 17 (c) and its equivalent points. The dots in red are Weyl points with positive chirality 1, while those in blue are Weyl points with negative chirality -1. Each Weyl point is coupled with the nearest one with opposite chirality.

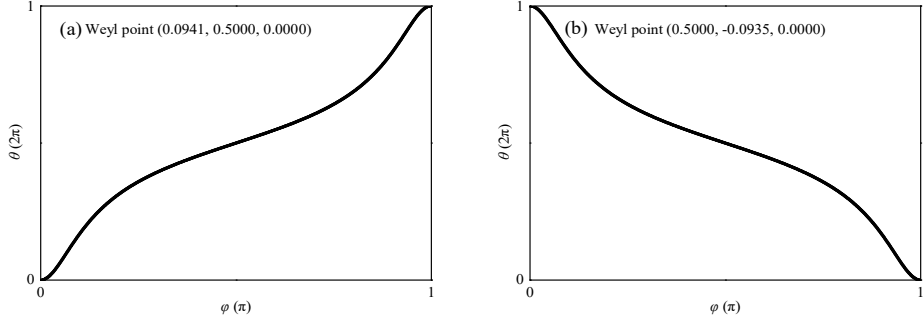

Figure 19. The Wannier charge center evolution for a couple Weyl points (0.0941, -0.5000, 0.0000) at (a) and (0.5000, -0.0935, 0.0000) at (b), with positive chirality 1, and negative chirality -1, respectively.

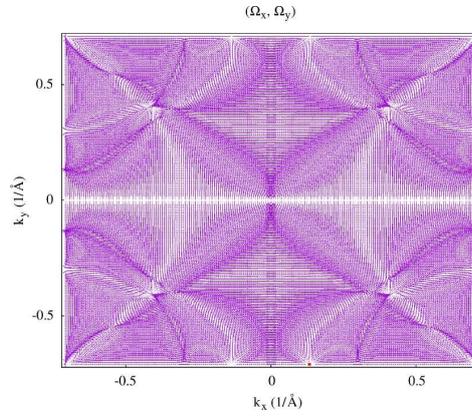

Figure 20. The Berry curvature distribution ($\Omega_x$, $\Omega_y$) in the $k_x$-$k_y$ plane with $k_z$=0 for the 6th phonon band. The source and sink can be found from the distribution. The red and blue dots represent the source and sink of Berry curvature distribution, illustrating positive and negative chirality. The boundary represents the $k_z$=0 plane of the Brillouin zone of $Al_3GaN_4$.

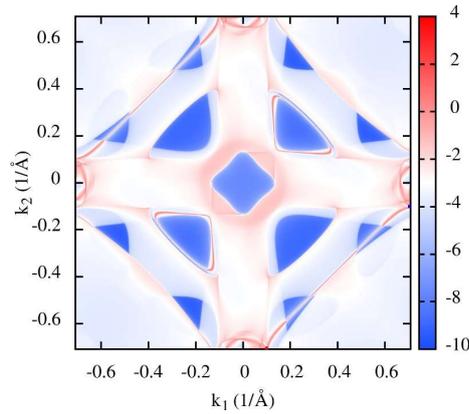

Figure 21. Weyl surface arcs. The projection of the iso-frequency plane of surface states on (001) surface at frequency 9.82 THz. The red and blue dots represent Weyl points with positive chirality 1

and negative chirality -1, respectively, which are connected by phonon surface arc. The boundary represents the Brillouin zone of the (001) surface of $Al_3GaN_4$.

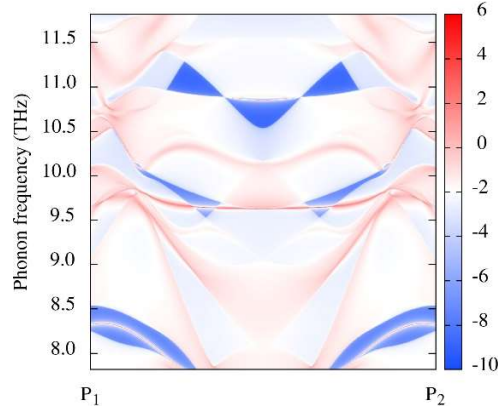

Figure 22. The LDOS for (001) surface along the selected path linked a couple of Weyl points. The single line in deeper red between a couple of Weyl points around 9.82 THz indicates the topological surface states, where $P_1$ and $P_2$ are on the extension line connecting two Weyl points.

## IV.  Conclusions and perspectives

In summary, we performed systematical analyses on the topological effects of phonons in wurtzite GaN with the aid of first-principle calculations and topological phonon theory by concentrating on Weyl point states and discussed the topological phonon phase transition in generalized GaN systems by introducing Al atoms, including AlN and AlGaN. The results show that wurtzite GaN is a trivial topological phonon system where no topological insulator phonon state and Weyl phonon state is found. However, by introducing Al atoms, Weyl phonon states are found in AlN and AlGaN of different structures, and they can be regarded as a topological phonon phase transition in AlGaN systems. By calculating topological charges, Wannier charge center evolutions, Berry curvature distributions, the projection of iso-frequency plane on the surface, and surface LDOS, two kinds of single Weyl phonon states in AlN, two kinds of single Weyl phonon states in $AlGaN_2$ with space group P3m1, and one single Weyl phonon state in cubic $Al_3GaN_4$ (which may not exist in reality due to NAC) are confirmed.

As typical materials in AlGaN/GaN heterostructures, wurtzite AlN and $AlGaN_2$ with space group P3m1 promise to host topological surface states without backscattering as they are confirmed to be topological phonon nontrivial systems with open Weyl surface arcs. Future research could directly examine these surface/interface states and the scattering characteristics between topological surface phonon states and trivial surface phonon states by molecular dynamic simulations. While AlN's thermal conductivity is relatively high, the contribution from topological surface phonon states may be insignificant. However, as mentioned in the first section of this work, interfacial thermal transport parallel to the AlGaN/GaN interface is critical for spreading of heat generated from near junction region. Hence, through effective design, the excellent transport properties of Weyl phonon states promise to improve the thermal management in information and power electronics.

**Acknowledgement**

We appreciated the help from Dr. Yizhou Liu (Weizmann Institute of Science, Israel) and Prof. Rui Wang (Chongqing University, China) in understanding the topological physics of phonons. This work was supported by the National Natural Science Foundation of China (Nos. 51825601 and U20A20301).

**Data Availability**

The data that support the findings of this study are available from the corresponding author upon reasonable request.